\documentclass[twocolumn,amsmath,amssymb,nofootinbib,superscriptaddress,longbibliography, floatfix]{revtex4-1}

\usepackage{graphicx}
\usepackage{amsmath}
\usepackage{amssymb}
\usepackage{xcolor}
\usepackage{hyperref}
\hypersetup{pdfstartview={FitH},pdfpagemode={UseNone}, breaklinks=true, colorlinks=true, linkcolor=blue, citecolor=blue, urlcolor=blue, bookmarksopen=true, pdfnewwindow=true}
\usepackage[all]{hypcap}
\usepackage[english]{babel}

\begin{document}

\preprint{AIP/123-QED}

\title{Wireless power transfer in magnetic resonance imaging at a higher-order mode of a birdcage coil}

\author{Oleg~I.~Burmistrov}
  \email{oleg.burmistrov@metalab.ifmo.ru}
\affiliation{School of Physics and Engineering, ITMO University, 49 Kronverksky pr., bldg. A, 197101 Saint Petersburg, Russia}

\author{Nikita~V.~Mikhailov}
\affiliation{School of Physics and Engineering, ITMO University, 49 Kronverksky pr., bldg. A, 197101 Saint Petersburg, Russia}

\author{Dmitriy~S.~Dashkevich}
\affiliation{School of Physics and Engineering, ITMO University, 49 Kronverksky pr., bldg. A, 197101 Saint Petersburg, Russia}

\author{Pavel~S.~Seregin}
\affiliation{School of Physics and Engineering, ITMO University, 49 Kronverksky pr., bldg. A, 197101 Saint Petersburg, Russia}

\author{Nikita~A.~Olekhno}
\affiliation{School of Physics and Engineering, ITMO University, 49 Kronverksky pr., bldg. A, 197101 Saint Petersburg, Russia}

\date{\today}

\begin{abstract}

Magnetic resonance imaging (MRI) is a crucial tool for medical visualization. In many cases, performing a scanning procedure requires the use of additional equipment, which can be powered by wires as well as via wireless power transfer (WPT) or wireless energy harvesting. In this Letter, we propose a novel scheme for WPT that uses a higher-order mode of the MRI scanner's birdcage coil for energy transmission. In contrast to the existing WPT solutions, our approach does not require additional transmitting coils. Compared to the energy harvesting, the proposed method allows supplying significantly more power. We perform numerical simulations demonstrating that one can use the fundamental mode of the birdcage coil to perform a scanning procedure while transmitting the energy to the receiver at a higher-order mode without any interference with the scanning signal or violation of safety constraints, as guaranteed by the mode structure of the birdcage. Also, we evaluate the specific absorption rate along with the energy transfer efficiency and verify our numerical model by a direct comparison with an experimental setup featuring a birdcage coil of a $1.5$~T MRI scanner.

\end{abstract}

\maketitle

\begin{figure}[b]
    \includegraphics[width=8.5cm]{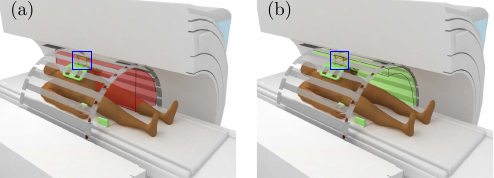}
    \caption{Artist’s view of WPT in MRI setup demonstrating the patient located at the table inside MR scanner bore, local coils and medical sensors (green), and the birdcage coil (gray). The birdcage creates a radio-frequency magnetic field either at (a) the fundamental mode (shaded with red) or (b) the higher-order mode (shaded with green). A receive system (shown with a blue frame) locates atop the patient.}
    \label{fig:idea}
\end{figure}

Magnetic resonance imaging (MRI)~\cite{kuperman2000magnetic} became one of the key medical visualization tools. MRI scanning procedures often rely on additional transmit and receive local coils, heart activity monitors, and other equipment placed near the patient inside an MR scanner bore~\cite{Nohava2020}, Fig.~\ref{fig:idea}(a). Typically, such devices are powered via special cables. However, such cables are bulky and may create imaging artifacts~\cite{Nohava2020}, reduce patient's comport, or even lead to thermal injuries~\cite{Dempsey2001thermal, dempsey2001investigation}. The other approach is based on wireless energy transmission methods, including two groups: wireless power transfer (WPT)~\cite{byron2017rf, byron2019, Ganti2019, ullah2022wireless} and wireless energy harvesting (WEH)~\cite{Ganti2021harv, Hofflin2013, riffe2007, Venkateswaran2020, SereginBurm2022Harv}. Despite applications of WPT~\cite{agbinya2022wireless, zhang2018wireless, song2017wireless} and WEH~\cite{bakir2018tunable, Hu2017, sharma2021electrically, Palazzi2018} are widely developed in other areas, they possess certain limitations.

The choice of an appropriate method depends on various factors, including the required transferred power level and MR scanner configuration. WPT allows transmitting higher power levels, but requires the presence of an additional transmitting antenna, which is typically not included in standard configurations of MR scanners. Besides, locating a transmitting WPT antenna inside a scanner bore is difficult due to the lack of free space and challenges in electromagnetic compatibility. On the other hand, WEH does not require an additional transmitting antenna inside a scanner bore, as it relies on converting a part of the excitation radio-frequency (RF) pulse energy at the fundamental mode frequency of a body coil. However, harvesting allows collecting only low power levels which are suitable for powering various simple sensors~\cite{SereginBurm2022Harv, Ganti2021harv}, but are unlikely to supply more complex equipment like local coils requiring more than $1$~W power~\cite{Nohava2020}. Moreover, the received power in this case depends on the applied pulse sequence~\cite{SereginBurm2022Harv}. Besides, WEH~\cite{Ganti2021harv, Hofflin2013, SereginBurm2022Harv} can considerably distort RF excitation field and decrease the quality of MR images compared to WPT~\cite{byron2019,  byron2017rf, Ganti2019} which uses the frequency different from the scanning one.

In this Letter, we propose a scheme for wireless power transfer in 1.5T MRI which does not require the use of additional transmitting antennas. In particular, we consider the modes of a birdcage coil~\cite{ahmad2020birdcage, vaughan2012rf} -- one of the most popular solutions for radio-frequency body coils in 1.5T setups -- and demonstrate that one can implement WPT at higher-order modes of the birdcage (which typically are not excited in MR scanners) while performing the scanning routine at the fundamental mode, Figure~\ref{fig:idea}(a,b), in contrast to previous schemes implementing WPT at the fundamental mode~\cite{rajendran2022birdcage}. The fundamental mode is characterized by a uniform distribution of the radio-frequency field $B_{1}^{+}$ inside the bore, which is used to excite nuclear magnetic resonance, Fig.~\ref{fig:idea}(a). As we demonstrate further, electromagnetic fields of higher-order modes are concentrated in the vicinity of the bore surface, i.e., in the area where potential receivers are placed, and are nearly absent in the central region where the patient locates, Fig.~\ref{fig:idea}(b). However, WPT at a higher-order mode can be implemented only within the transmit mode of the scanning procedure because a body coil is typically detuned while receiving $B_{1}^{-}$ field if a local coil is used during the scan~\cite{vaughan2012rf}.

\begin{figure}[tb]
    \includegraphics[width=8.5cm]{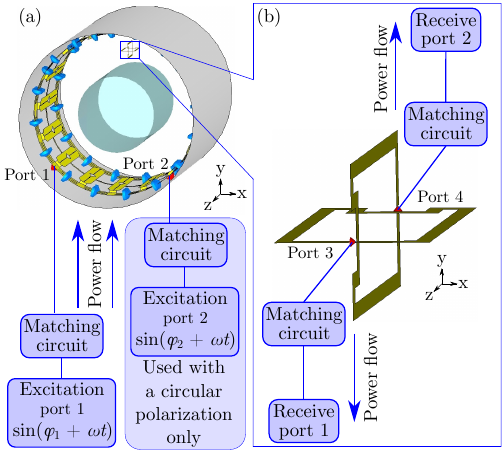}
    \caption{(a) Schematics of WPT at a higher-order mode of a birdcage coil. The numerical model includes a birdcage coil (yellow stripes), capacitors (blue), ports (red), an RF screen (olive cylinder), a phantom (blue transparent cylinder), and a receive system (shown with the blue frame). (b) Receive system consisting of two orthogonal loop coils allowing to convert horizontal and vertical polarizations of the magnetic field. Blue arrows indicate the direction of power flows.}
    \label{fig:setup}
\end{figure}


A typical structure of a high-pass birdcage coil is shown in Fig.~\ref{fig:setup}(a). We consider the coil consisting of $N=16$ copper stripes placed at the side surface of a cylinder and connected by capacitors. The complete schematics is given in Supplementary Materials. The diameter of the coil is $636$~mm, and the length is $575$~mm. Such parameters correspond to the geometry and sizes of Siemens Avanto 1.5T body coil. The numerical model includes two ports [red markers in Fig.~\ref{fig:setup}(a)] connecting the first and the fifth stripes of the birdcage with the surrounding electromagnetic shield, and the phantom which emulates the load created by a patient inside the scanner, Fig.~\ref{fig:setup}(a). The diameter of the phantom is $300$~mm, the length is $500$~mm, the permittivity is $\varepsilon=80$, and the conductivity is $1$~Sm/m, corresponding to standard MRI Siemens phantom at 1.5T. Finally, the model includes a WPT receive system composed of two orthogonal loop coils, shown in Fig.~\ref{fig:setup}(b). In the numerical model, we tune the resonance frequency to the value $61.67$~MHz measured in our experimental setup, which is close to the Larmor frequency for 1.5T MRI ($63.55$~MHz). All numerical simulations were performed in CST Microwave Studio 2022.

\begin{figure*}[t]
    \includegraphics[width=17cm]{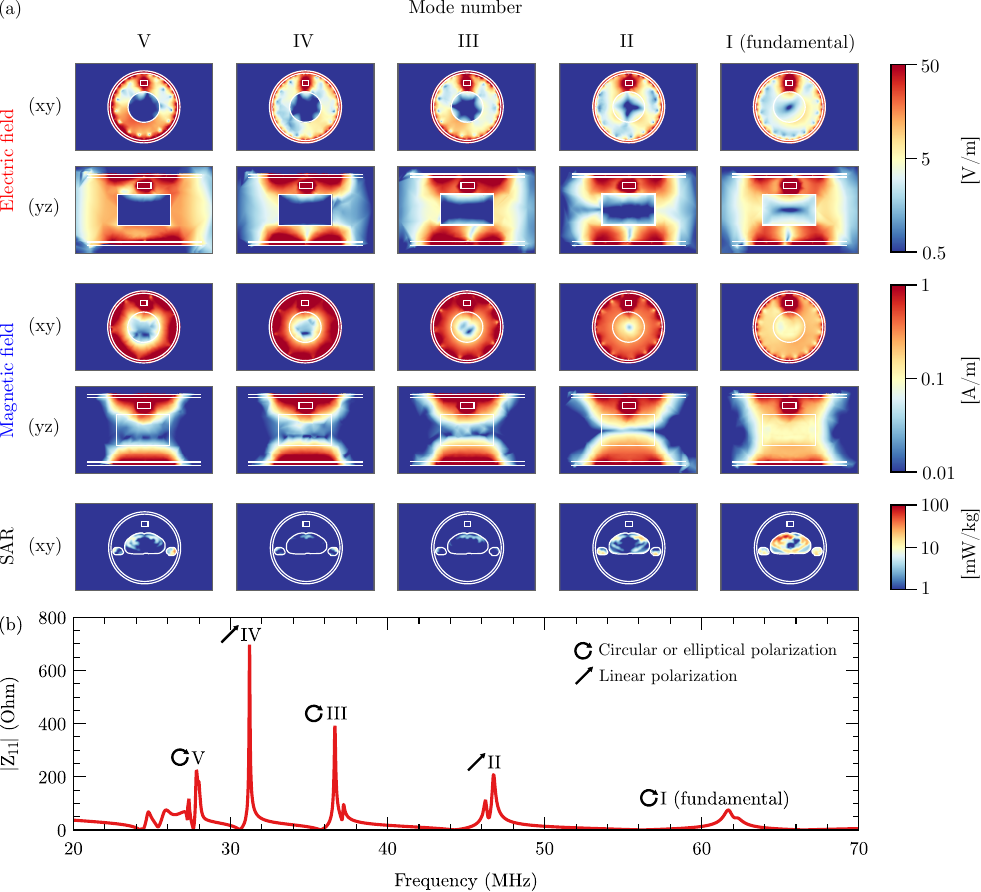}
    \caption{(a) Numerical simulations of the electric and magnetic fields and specific absorption rate (SAR) for different modes of the birdcage coil in the axial (xy) and sagittal (yz) planes. The modes are enumerated from (I) to (V) starting with the fundamental mode which has the highest frequency. (b) Impedance spectrum ($Z_{11}$-parameter) at the first port of the birdcage coil Figure~\ref{fig:setup}(a) demonstrating characteristic resonances of different modes. The birdcage coil is considered with a cylindrical phantom (for electric and magnetic fields) and with a human voxel model (for SAR) shown with white contours. The simulations in (b) are performed without the receive system.}
    \label{fig:simulations}
\end{figure*}

The receive system consists of two identical orthogonal loop coils, the vertical and the horizontal ones, Fig.~\ref{fig:setup}(b). The loop coils are rectangular copper frames with dimensions of $100 \times 50$~mm and the width of $5$~mm. The center of the receive system is located at the height of $225$~mm above the center of birdcage, Fig.~\ref{fig:setup}(a). Each loop in the numerical model is supplied with a single port [red markers in Fig.~\ref{fig:setup}]. Such a system allows simultaneously receiving vertically and horizontally polarized linear components of the RF magnetic field, and, as a result, efficiently convert circularly polarized fields~\cite{SereginBurm2022Harv}. 

\begin{figure}[tbp]
    \includegraphics[width=8.5cm]{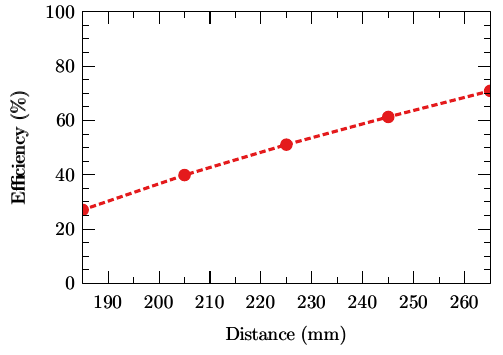}
    \caption{The dependence of power transfer efficiency Eq.~\eqref{eq:efficiency_calc} on the distance between the centers of the birdcage and the receive system obtained numerically. The distance is varied between $185$~mm (corresponding to the separation of $10$~mm between the phantom and the lower edge of the receive system) and $265$~mm (corresponding to the separation of $10$~mm between the bore and the upper edge of the receive system) in the geometry of Fig.~\ref{fig:setup}(a). The distances $185$~mm, $205$~mm, $225$~mm, $245$~mm, and $265$~mm are considered.}
    \label{fig:efficiency}
\end{figure}

Schematic block within the numerical model includes four ports: two excitation ports linked to the birdcage coil [Fig.~\ref{fig:setup}(a)] and two receive ports [Fig.~\ref{fig:setup}(b)]. All four schematic ports are connected to the respective ports in models of the birdcage and receive system through L-section matching circuits. All lumped elements in the model feature equivalent serial resistance of $0.1$~Ohm. In most cases, receive and excitation ports are matched to $Z = 50$~Ohms, except when calculating the resonance frequencies of the birdcage and comparing the numerical results for WPT with the experimental measurements.

\begin{figure*}[tbp]
    \includegraphics[width=17cm]{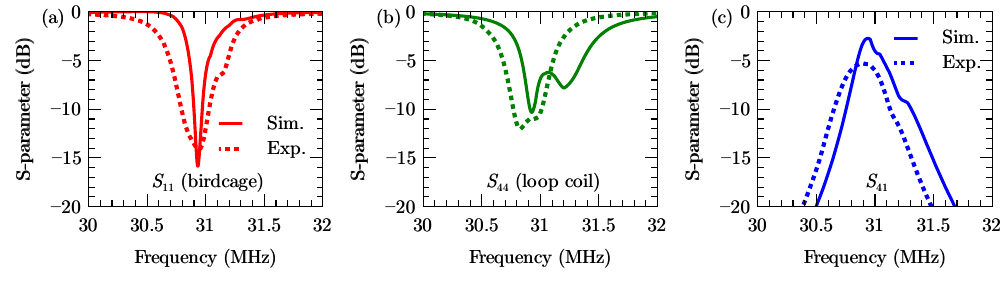}
    \caption{The comparison of experimentally measured (dashed lines) and numerically simulated (solid lines) $S$-parameters. (a) $S_{11}$-parameter for the first port of the birdcage. (b) $S_{44}$-parameter for the vertical loop antenna. (c) $S_{41}$-parameter (transmission coefficient) between the first port of the birdcage and the vertical loop antenna.}
    \label{fig:experiment}
\end{figure*}


In our numerical simulations, we start with calculating the $Z_{11}$-parameter of the birdcage with a body phantom for different excitation frequencies without any receiving system in the range from $20$~MHz to $70$~MHz and observe a series of resonances shown in Fig.~\ref{fig:simulations}(b). The peak at $61.67$~MHz corresponds to the circularly polarized fundamental mode of the birdcage used for scanning, while peaks at lower frequencies are associated with higher-order modes. Note that while the Modes (III) and (V) share the circular polarization, Modes (II) and (IV) are linearly polarized. Due to the strong coupling between the two birdcage ports for these linearly polarized modes, we use a single port for their excitation. All used excitation ports are matched to $50$~Ohms.

Next, we evaluate the distributions of electric field, magnetic field, and specific absorption rate (SAR) for the birdcage excitation at the respective mode frequencies from Fig.~\ref{fig:simulations}(b), but now introducing the receiving system to the model, Fig.~\ref{fig:simulations}(a). It is seen that the magnetic fields in the axial ($H_{xy}$) and sagittal ($H_{yz}$) planes for the fundamental mode remain uniform in the phantom volume, as required for the scanning procedure, and the presence of the receive system introduces only slight distortions in its vicinity. However, for higher-order modes the situation is completely different: the higher the mode number, the lower is the magnetic field in the phantom. In particular, for Modes (III-V), the magnetic field in the phantom nearly vanishes. Moreover, the frequencies of these modes significantly differ from the fundamental frequency. As a result, transmitting power at these modes will not affect the scanning procedure or create any imaging artifacts. A similar behavior is observed for the electric fields: the respective quantities vanish in the phantom area for higher-order modes despite a pronounced electric field concentration observed at the receive system, highlighting the safety of the proposed WPT scheme.

Next, we evaluate the WPT efficiency at different higher-order modes, see Table~\ref{tabular:power}. We obtain voltages and currents at Excitation ports~1,2 and Receive ports~1,2 and evaluate the corresponding values of power $P_{\rm tx1, tx2}$ transmitted by the birdcage coil ports 1,2 and $P_{\rm rx1, rx2}$ received by the horizontal and vertical loop coils of the receive system, respectively. Then, we define the efficiency $\eta$ as
\begin{equation}
   \eta = \frac{P_{\rm rx1} + P_{\rm rx2}}{P_{\rm tx1} + P_{\rm tx2}}.
   \label{eq:efficiency_calc}
\end{equation}
In the case of linear polarization, the birdcage is excited with a single Excitation port~1 by supplying $1000$~mW, while for circular polarization both excitation ports are used simultaneously powered with $500$~mW each. As a result, for the modes with linear polarization $P_{\rm tx2}=0$. The voltage from Excitation port~2 attains an additional phase of $-90^\circ$. The lowest efficiency of $20.7\%$ is displayed by the fundamental mode, as its magnetic field is uniformly distributed in the bore volume and has low values near its edges where the receive system is located. Then, for higher-order modes (II) to (IV) the efficiency monotonically grows from $38\%$ for Mode (II) and $49.9\%$ for Mode (III) to $51\%$ for Mode (IV). Such a growth is associated with increased magnetic field localization at the bore edge (i.e., in the vicinity of the receive system), as seen in Fig.~\ref{fig:simulations}(a). Then, for Mode (V) the efficiency drops to $26.3\%$ due to a substantial increase in the electric fields and an associated decrease in the magnetic field for a fixed power transmitted by the birdcage.

\begin{table}[tbp]
   \caption{\label{tabular:power} Transmitted and received power levels for different modes of the birdcage coil obtained in numerical simulations. The frequency of the corresponding mode is $f$, $P_{\rm tx1}$ is the power transferred through the first schematic port, $P_{\rm tx2}$ is the power transferred through the second schematic port ("N/A" for linear polarization), $P_{\rm rx1}$ is the power received by the horizontal loop, and $P_{\rm rx2}$ is the power received by the vertical loop. The power transfer efficiency $\eta$ is calculated by Eq.~\eqref{eq:efficiency_calc}.}
   \begin{ruledtabular}
   \begin{tabular}{c c c c c c c}
   Mode & $f$,~MHz & $P_{\rm tx1}$,~mW & $P_{\rm tx2}$,~mW  & $P_{\rm rx1}$,~mW & $P_{\rm rx2}$,~mW & $\eta$,~\% \\ \hline
   (I) & 61.67 &  500 & 500 &  34 & 173 & 20.7   \\
   (II) & 46.76 & 1000 &  N/A  & 299 &  81 & 38   \\
   (III) & 36.65 &  500 & 500 &  117 & 382 & 49.9 \\
   (IV) & 31.21 & 1000 &  N/A  &  78 & 432 & 51 \\
   (V) & 27.88 &  500 & 500 &  216 & 47 & 26.3 \\
   \end{tabular}
   \end{ruledtabular}
\end{table}

Finally, we numerically evaluate the local SAR (characterizing the local heating of the patient), see Fig.~\ref{fig:simulations}(a), and the whole-body SAR (characterizing the entire body heating). For the studies of SAR, the cylindrical phantom is replaced with Gustav voxel human model which includes all tissues and organs and is a part of CST Bio Models Library Extension. The maximum values of the local ($217$~mW/kg) and the whole-body ($5.79$~mW/kg) SAR correspond to the fundamental mode, as electric and magnetic RF fields penetrate deep into the scanning object, Fig.~\ref{fig:simulations}(a). The minimal values of the local SAR ($7.8$~mW/kg) and the whole-body SAR ($0.299$~mW/kg) correspond to Mode (III) and Mode (IV), respectively. Since MRI possesses the limitation of $2$~W/kg for the local SAR~\cite{sar_standart}, the maximum achievable power at the receive system is $128$~W for Mode (III) considering the system's center located at the height of $225$~mm above the birdcage center. Such an amount allows fully supplying common types of local receive coils~\cite{Nohava2020}. The maximum received power levels are $72$~W at Mode (IV), $16$~W at Mode (II), $5$~W at Mode (V), and $2$~W at the fundamental mode. Technical limitations for WPT with higher-order modes associated with an electromagnetic compatibility with local coils are discussed in Supplementary Materials.

For further studies, we select Mode (IV) considering its lower SAR compared to Modes (II) and (V) and the highest power transfer efficiency among all the considered modes. Figure~\ref{fig:efficiency} demonstrates the dependency of power transfer efficiency for Mode (IV) on the distance between the receiving system center and the birdcage center. It is seen that the efficiency monotonically increases with the distance in accordance with the similar growth in the magnetic field in the vicinity of the bore edge, Fig.~\ref{fig:simulations}(a).

To support our numerical findings, we realize WPT at Mode (IV) with the help of experimental setup which includes a birdcage coil from Siemens Avanto 1.5~T scanner, a matching circuit for the birdcage, an RF screen made of a copper grid, a phantom, a receive system with a matching circuit, and measurement equipment. Prior to the implementation of matching circuits for Mode (IV), they were simulated with CST Microwave Studio 2022. The values of capacitances and resistances in the numerical model of the birdcage coil were obtained by measuring the components used in the experimental setup.

Figure~\ref{fig:experiment} shows the S-parameters characterizing the reflection of the signal for the birdcage coil port $S_{11}$ [Fig.~\ref{fig:experiment}(a)], the receive system port $S_{44}$ [Fig.~\ref{fig:experiment}(b)], and the transmission coefficient between the birdcage and the receive system  $S_{41}$ [Fig.~\ref{fig:experiment}(c)]. It is seen that the numerical and experimental values of S-parameters agree well, up to some broadening and lower amplitude of the experimental peak in Fig.~\ref{fig:experiment}(a) which may be associated with higher losses in the experimental setup compared to the numerical model. Also, the positions of the receive system resonances in Fig.~\ref{fig:experiment}(b) appear shifted due to high sensitivity of their positions to the values of tuning capacitors which may differ in numerical simulation and the experimental setup. These results highlight the possibility of implementing such WPT scheme in a real MRI setup.


In conclusion, we demonstrated that one can implement wireless power transfer with the help of a higher-order mode of a birdcage coil. The main advantage of such an approach is the absence of the need to place an additional transmitting WPT coil and associated power supply cables inside the MRI bore, as such metallic objects can introduce imaging artifacts. Also, the proposed solution allows lowering the price for WPT system. We have studied numerically the structure of higher-order modes for a typical birdcage coil and showed that it should not introduce any imaging artifacts, evaluated the power transfer efficiency which may reach $51\%$ and the maximal transmitted power which is $128$~W, and demonstrated that such an approach is safe for a patient by calculating the specific absorption rate. However, the WPT efficiency considerably depends on the location of the receive system, and the power transfer is performed only during the transmission mode which typically takes just $0.1\%$ of the overall working time, thus limiting the amount of transmitted power. The future directions for developing the proposed design include incorporating the receive system into a local coil and performing an experiment in the MRI scanner during the scanning procedure.


See the supplementary material for detailed descriptions of the birdcage coil schematics and the experimental setup design, the receive system structure, matching circuits schematics, local coil electromagnetic compatibility engineering, and the study of the fundamental mode frequency dependence on the screen diameter.

The authors acknowledge fruitful discussions with Prof. Stanislav Glybovski. This work was supported by the Russian Science Foundation (Project No.21-79-30038).

\section*{Author Declarations}

\subsection*{Conflict of Interest}
The authors have no conflicts to disclose.

\subsection*{Author Contributions}

\textbf{Oleg~I.~Burmistrov:} Conceptualization (supporting); Investigation (lead); Resources (lead); Software (lead); Supervision (supporting); Visualization (lead); Writing -- original draft (supporting).
\textbf{Nikita~V.~Mikhailov:} Investigation (supporting); Software (supporting).
\textbf{Dmitriy~S.~Dashkevich:} Investigation (supporting); Resources (supporting).
\textbf{Pavel~S.~Seregin:} Conceptualization (lead); Investigation (supporting); Project Administration (supporting); Supervision (supporting); Writing -- original draft (supporting).
\textbf{Nikita~A.~Olekhno:} Conceptualization (supporting); Project Administration (lead); Supervision (lead); Visualization (supporting); Writing -- original draft (lead).

\section*{Data Availability}
The data that support the findings of this study are available from the corresponding author upon reasonable request.

\end{document}



\title{Supplementary Material\\Wireless power transfer in magnetic resonance imaging at a higher-order mode of a birdcage coil}

\author{Oleg~I.~Burmistrov}
\email{oleg.burmistrov@metalab.ifmo.ru}
\affiliation{School of Physics and Engineering, ITMO University, 49 Kronverksky pr., bldg. A, 197101 Saint Petersburg, Russia}

\author{Nikita~V.~Mikhailov}
\affiliation{School of Physics and Engineering, ITMO University, 49 Kronverksky pr., bldg. A, 197101 Saint Petersburg, Russia}

\author{Dmitriy~S.~Dashkevich}
\affiliation{School of Physics and Engineering, ITMO University, 49 Kronverksky pr., bldg. A, 197101 Saint Petersburg, Russia}

\author{Pavel~S.~Seregin}
\affiliation{School of Physics and Engineering, ITMO University, 49 Kronverksky pr., bldg. A, 197101 Saint Petersburg, Russia}

\author{Nikita~A.~Olekhno}
\affiliation{School of Physics and Engineering, ITMO University, 49 Kronverksky pr., bldg. A, 197101 Saint Petersburg, Russia}

\date{\today}

\maketitle

\begin{spacing}{1.5}

\tableofcontents

\section*{Supplementary Note 1. The birdcage coil schematics and the experimental setup}

In this Section, we describe the setup used for experimental studies described in the main text. The setup includes the birdcage coil from Siemens Avanto 1.5~T MRI scanner, a matching circuit for the birdcage, an RF screen made of a copper grid, a phantom, a receive system with a matching circuit, and measurement equipment.

\begin{figure*}[tbp]
    \includegraphics[width=17cm]{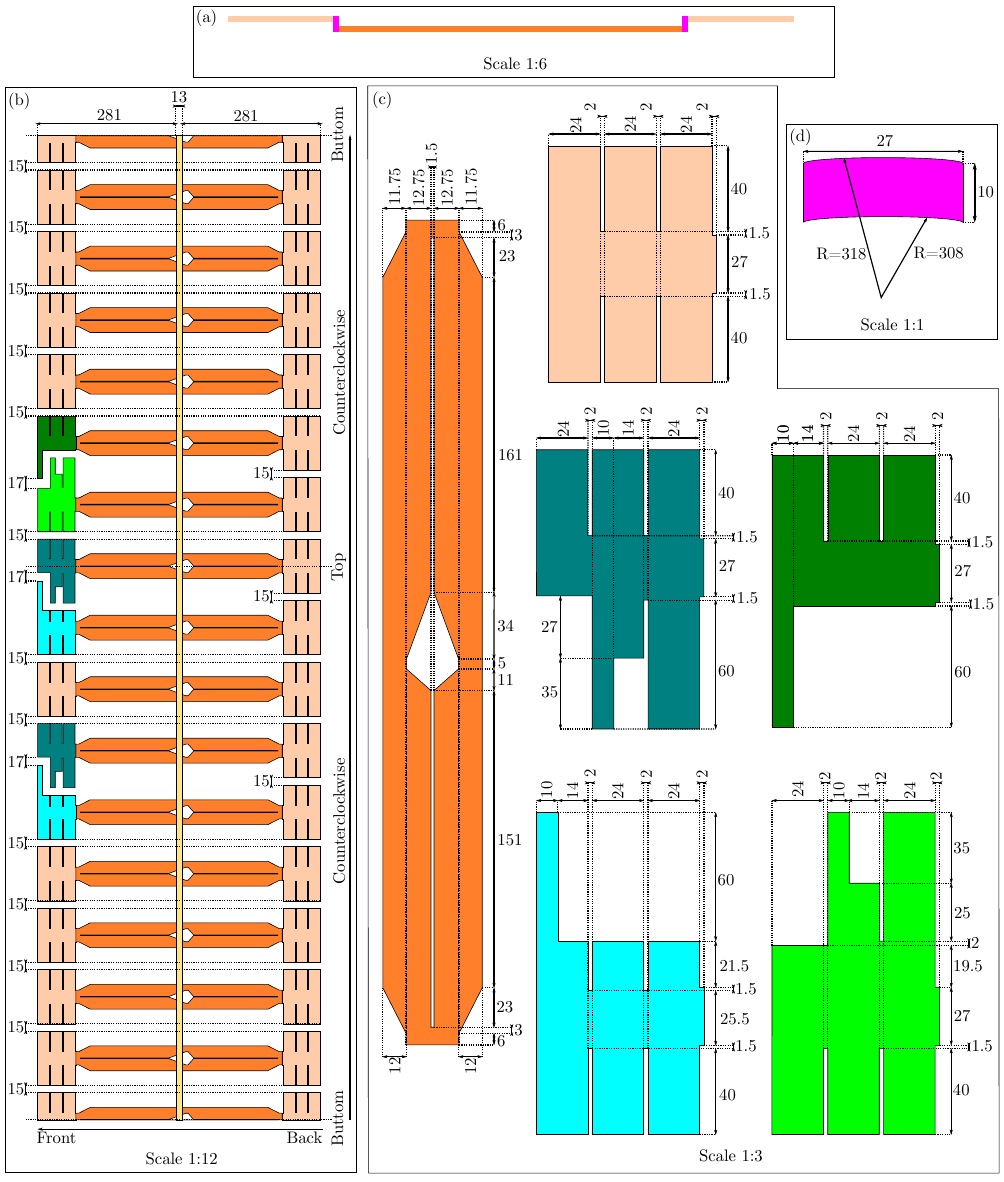}
    \caption{Geometry of a birdcage used in numerical simulations and experimental studies. (a) Side view of the unwrapped birdcage geometry showing the connection between the copper stripes and patches. (b) Top view of the unwrapped birdcage geometry. (c) In-plane sizes of the copper stripes (orange) and different patches (olive, navy, cyan, green, and peach). (d) In-plane sizes of the patches shown in purple. The distances from the center of the wrapped birdcage to the bottom $R=308$~mm and the top $R=318$~mm edges of the patch are shown. The scale for all panels is displayed at the bottom of the respective panels.}
    \label{fig:geometry_of_bc}
\end{figure*}

\begin{figure*}[tbp]
    \includegraphics[width=17cm]{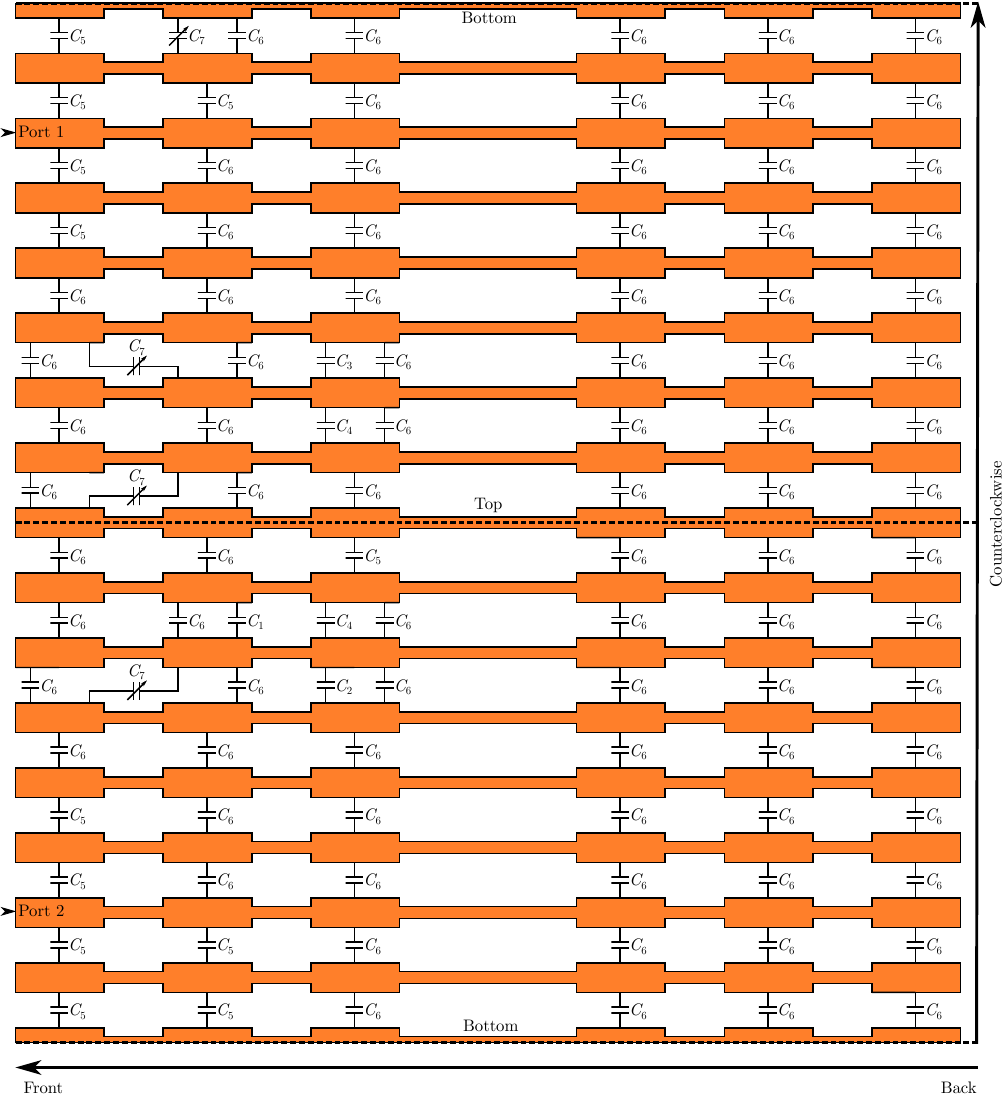}
    \caption{Capacitors and power ports in the numerical model. $C_1$ = $1$~pF, $C_2$ = $2.2$~pF, $C_3$ = $5.6$~pF, $C_4$ = $12$~pF, $C_5$ = $39$~pF, $C_6$ = $47$~pF, and $C_7$ is the variable capacitor CPC-MT~8/30~pF.}
    \label{fig:scheme}
\end{figure*}

In our numerical simulations and experimental studies, we consider a coil with $N=16$ copper stripes of the width $50.5$~mm [Fig.~\ref{fig:geometry_of_bc}] and the following capacitors: $1$~pF, $2.2$~pF, $5.6$~pF, $12$~pF, $39$~pF, $47$~pF, and a variable capacitor, Fig.~\ref{fig:scheme}. In the numerical model, the capacitors are implemented as lumped elements with $R=0.1$~Ohm serial resistance (measured in our experimental setup with Keysight E4991B Impedance Analyzer). The diameter of the coil is $636$~mm, and the length is $575$~mm.

Figure~\ref{fig:geometry_of_bc} shows the unwrapped pattern of the birdcage coil. The coil consists of copper stripes (orange) and five types of patches (olive, green, navy, cyan, and peach) having different geometry and attached to the ends of the stripes via elements shown with purple. The yellow ring is not connected to the copper stripes and is used for opening the diodes of the detuning system with an additional potential. Figures~\ref{fig:geometry_of_bc}(a,b) demonstrate the connection geometry, while Figs.~\ref{fig:geometry_of_bc}(c,d) include the in-plane sizes of the elements. Such dimensions of the setup are used in numerical studies as well.

Several modifications were made to the body coil. First, the detuning system was removed as the coil is used in the excitation regime only. Second, the diodes within the coil were removed and replaced by copper wires with the diameter of $1$~mm, since these diodes are needed only in the receive mode to decouple the body coil and local coils. Also, $5.6$~nF capacitors were replaced by metallic stripes as they are equivalent to short circuit in the frequency region of interest. In the MRI scanner, such capacitors are applied to damp the eddy currents created by the gradient coils which are absent in our setup. Besides that, both power ports of the birdcage are replaced by SMA female ports which are connected to the RF screen and to the first and the fifth legs of the body coil, respectively. In the experimental setup, such ports are used to connect matching circuits. Finally, the original variable capacitors were replaced by 8/30~pF KPK-MN capacitors to increase the tuning convenience. A variable capacitor was added between the second and the third legs of the birdcage for additional studies. However, in the final setup, the capacitor was set to the minimal value $C \approx 2$~pF.

Figure~\ref{fig:scheme} shows the schematics of capacitor placement between the metallic parts and power ports connection. Capacitors $C_{1}..C_{6}$ are high voltage ($7$~kV breakdown voltage) and $C_{7}$ is a variable capacitor. The circuit schematics of Fig.~\ref{fig:scheme} is used both in numerical simulations and experimental studies.

The RF screen is made from a copper mesh with $1 \times 2$~mm cells fixed by an adhesive tape. The screen is grounded and insulated from the body coil by polyurethane foam featuring low permittivity $\varepsilon \sim 1$, which can be omitted in numerical simulations. The fundamental mode frequency is $63.55$~MHz, which slightly differs from the value in numerical simulations ($61.67$~MHz).

The phantom used in our experimental setup is 2377425-3 Body Array Phantom by General Electric located at the birdcage isocenter. In numerical simulations, we use the values of permittivity $\varepsilon=80$ and conductivity $\sigma=1$~Sm/m correspond to the liquid from analogous standard MRI Siemens phantom plastic bottle PN~5512608. Such a phantom is more convenient for measurements, and we obtained these values experimentally with the help of SPEAG Dielectric Assessment Kit~12 and Planar~S5048 vector network analyzer.

\begin{singlespace}
\end{singlespace}

In CST Microwave Studio, root mean squared voltages and currents at the four schematic ports are evaluated in the entire frequency range from $20$~MHz to $70$~MHz with the help of AC-task. Moreover, this function allows finding field polarizations, combining the results evaluated in the schematic block with the 3D Finite Element Method (FEM) models, and evaluating the electromagnetic fields, excitation voltages, and loads in the ports with the proper matching circuits. The SAR distributions are simulated with a time domain solver based on modified Finite Difference Time Domain (FDTD) method.

\section*{Supplementary Note 2. The receive system}

\begin{figure*}[tbp]
    \includegraphics[width=17cm]{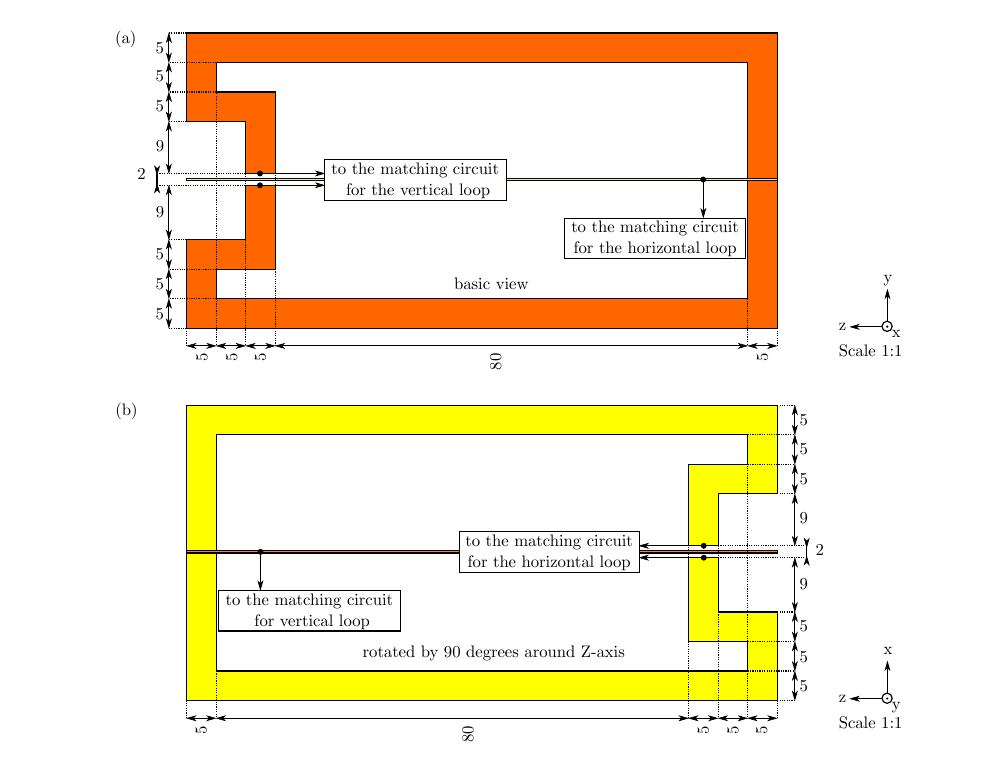}
    \caption{Geometry of the receive system used in numerical simulations and experimental studies in (a) the sagittal (yz) and (b) the coronal (xz) planes of view.}
    \label{fig:geometry_of_receive_system}
\end{figure*}

The receive system consists of two orthogonal copper loop coils having identical geometries, Fig.~\ref{fig:geometry_of_receive_system}. The in-plane sizes of each individual coil are $50 \times 100$~mm, and the thickness of copper foil is $35~\mu{\rm m}$. In numerical simulations, each coil has a single port including two contact points: one connecting it to the corresponding matching circuit and the other one connecting the coil to the ground. Each coil converts the respective component of the magnetic field, thus allowing to efficiently receive a circularly polarized magnetic field characteristic of a majority of MR scanners. Such a receive system develops our previous design used for wireless energy harvesting in MRI~\cite{SereginBurm2022Harv}.

\section*{Supplementary Note 3. Matching circuits for numerical simulations and experimental studies}

\begin{figure*}[tbp]
    \centering
    \includegraphics[width=10cm]{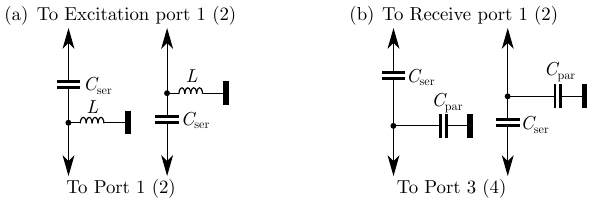}
    \caption{(a) Matching circuits for Excitation ports~1,2. (b) Matching circuits for Receive ports~1,2.}
    \label{fig:matching_circuits}
\end{figure*}

The excitation ports [Port~1 and Port~2 at Fig.~2(a) in the main text] of the birdcage are always matched with a combination of an inductance coil $L$ and a capacitor $C_{\rm ser}$ [Fig.~\ref{fig:matching_circuits}(a)], while for the receive ports [Port~3 and Port~4 at Fig.~2(b) in the main text] a combination of two capacitors $C_{\rm ser}$ and $C_{\rm par}$ is used [Fig.~\ref{fig:matching_circuits}(b)]. For all numerical simulations, Excitation ports~1,2 and Receive ports~1,2 are matched to the impedance ${{\rm Re}\{Z\} = 50 \pm 1}$~Ohm, ${\rm Im}\{Z\} = 0 \pm 2j$~Ohm. However, there are two exceptions: when determining the resonance frequencies of birdcage modes (no matching circuit is applied) and during the verification of the numerical model by a comparison with the experimental setup (the matching circuit consists only of capacitors without fine tuning of the matching parameters). 

Resonance frequencies of the birdcage coil and the values of components for the matching circuits were calculated with the help of S-parameter study in the schematic block upon setting the impedance of Excitation ports~1,2 and Receive ports~1,2 to $50$~Ohm. Since there are several dozens of different values of capacitors and inductors for various matching circuits considering different matching for the evaluation of electric and magnetic fields and SAR, as well as for different excitation polarizations and different working frequencies of the modes, we do not include a complete list of the components. The values of matching capacitors and inductors are in the range from $10$~pF to $1$~nF and from $10$~nH to $10~\mu{\rm H}$, respectively.

\section*{Supplementary Note 4. Local coil electromagnetic compatibility engineering}

\begin{figure*}[tbp]
    \includegraphics[width=17cm]{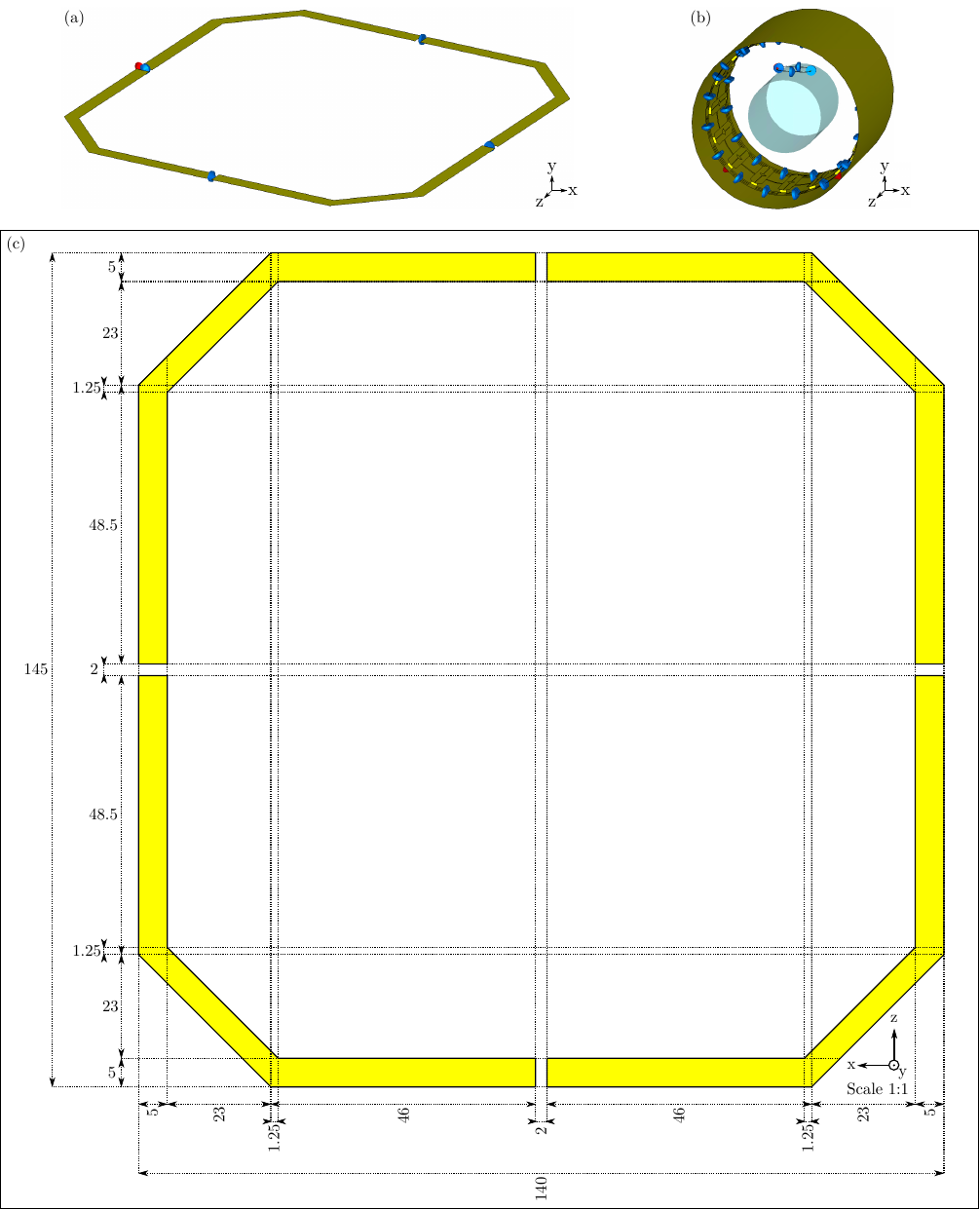}
    \caption{The numerical model for electromagnetic compatibility simulations. (a) 3D FEM model of the loop coil from Siemens Spine Array for 1.5T MRI. Capacitors are shown with blue discs, the red marker denotes the port. (b) In simulations, the loop coil locates at the height of $10$~mm upon the phantom surface within the birdcage coil and the screen. (c) In-plane sizes of the loop coil.}
    \label{fig:local_coil_geometry}
\end{figure*}

\begin{figure*}[tbp]
    \includegraphics[width=17cm]{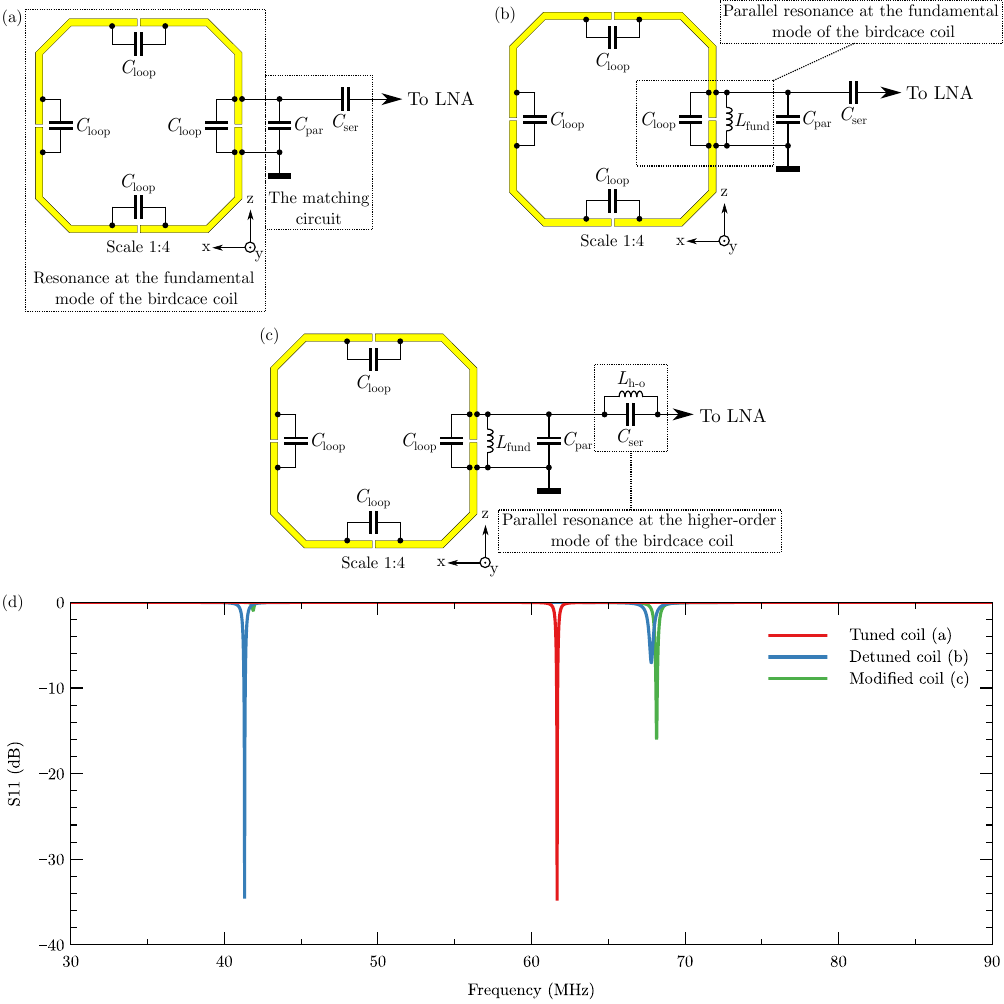}
    \caption{The schematics of the local coil model used in electromagnetic compatibility simulations. The equivalent serial resistance (ESR) of all lumped elements is $0.1$~Ohm. The impedance of low noise amplifier (LNA) is $50$~Ohm. (a) The local coil is tuned to the fundamental mode of the birdcage coil. (b) The detuning inductor ($L_{\rm fund}$) is added in parallel to $C_{\rm loop} = 57$~pF in the circuit shown in panel (a). (c) The modified detuning circuit for WPT at the birdcage coil higher-order mode. The detuning inductor $L_{\rm h-o}$ is added in parallel to $C_{\rm ser}$ in the circuit shown in (b). (d) $S_{\rm 11}$ parameter before LNA (impedance $50$~Ohm) for (a) the tuned coil, (b) the detuned coil, and (c) the modified coil. The lumped element values are the following: $C_{\rm ser} = 6.8$~pF, $C_{\rm par} = 24.4$~pF, $L_{\rm fund} = 110$~nH, $L_{\rm h-o} = 2.77~\mu{\rm H}$.}
    \label{fig:local_coil_circuits}
\end{figure*}

In this Section, we address possible limitations related to the electromagnetic compatibility of the WPT system and a local coil. We perform numerical simulations for Siemens loop coil. The corresponding model includes copper stripes, four capacitors $C_{\rm loop}$ = $57$~pF, and the port, Fig.~\ref{fig:local_coil_geometry}(a). The local coil is positioned at $10$~mm above the phantom surface [Fig.~\ref{fig:local_coil_geometry}(b)] along the Y-axis. The birdcage coil with the screen, the phantom, and the local coil have a common center at the X- and Z-axes. The sizes of the local coil are shown in Fig.~\ref{fig:local_coil_geometry}(c).

The schematic block includes L-shape matching circuits for the birdcage coil [$L$ and $C_{\rm ser}$, Fig.~\ref{fig:matching_circuits}(a)] and the local coil [$C_{\rm ser}$ and $C_{\rm par}$, Fig.~\ref{fig:local_coil_circuits}(a)], the detuning system in the form of a parallel LC-circuit for the resonance frequency at the fundamental [$L_{\rm fund}$ and $C_{\rm loop}$, Fig.~\ref{fig:local_coil_circuits}(b)] and the higher-order [$L_{\rm h-o}$ and $C_{\rm ser}$, Fig.~\ref{fig:local_coil_circuits}(c)] birdcage coil modes, respectively. A parallel LC-circuit has a high impedance at the resonance frequency, and, as a result, the signal (power flow) cannot pass it through $50$~Ohm transfer line. The local coil with the matching circuit have a resonance at the fundamental mode [Fig.~\ref{fig:local_coil_circuits}(d)]. If the local coil is detuned, it has two resonances: at $f=40.38$~MHz (which can be close to the frequency of a higher-order mode) and $f=66.36$~MHz [Fig.~\ref{fig:local_coil_circuits}(d)]. For such a case, we obtain the input power for the birdcage coil of $4.7$~W for Mode (II), $80$~W for Mode (III), and $810$~W for Mode (IV) if the local coil is detuned, Fig.~\ref{fig:local_coil_circuits}(b). The local coil port in Fig.~\ref{fig:local_coil_circuits}(a-c) emulates a preamplifier -- a low noise amplifier (LNA) which breaks out at the input voltage threshold of $0.7$~V.

Next, we add the inductor $L_{\rm h-o}$ to the serial capacitance $C_{\rm ser}$ in the matching circuit of Fig.~\ref{fig:local_coil_circuits}(c). As a result, the input power of the birdcage coil with a detuned local coil can reach $147$~MW at the higher-order mode. It is seen the parallel LC-circuit reduces the $S_{11}$ amplitude near the higher-order resonance frequency, Fig.~\ref{fig:local_coil_circuits}(d), thus preventing the possible coupling of the local coil with a higher-order mode of the birdcage and the WPT system.

\section*{Supplementary Note 5. The dependence of the fundamental mode frequency on the screen diameter}

\begin{figure*}[tbp]
    \includegraphics[width=17cm]{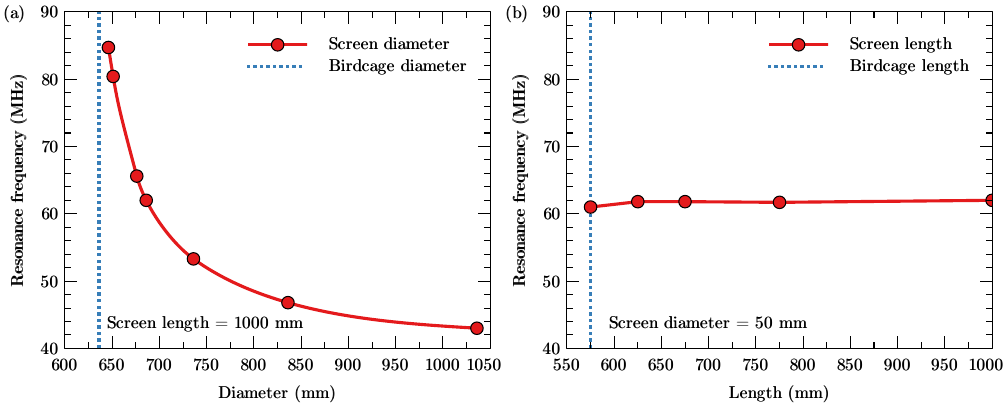}
    \caption{The dependence of the resonance frequency on the screen diameter (a) and the screen length (b). Blue dotted lines show the birdcage coil sizes.}
    \label{fig:screen}
\end{figure*}

The cylindrical copper RF screen [Fig.~2(a) in the main text] has the diameter of $685.6$~mm, and its length is $1000$~mm. Geometrical parameters of the screen considerably affect the resonance frequency of a birdcage coil. Changing the screen diameter by $10$~mm shifts the body coil resonance frequency by $1$~MHz, Fig.~\ref{fig:screen}(a), while the screen length has a minimal effect, Fig.~\ref{fig:screen}(b). In the numerical model, we set the screen diameter $685.6$~mm to tune the resonance frequency to the value $61.67$~MHz measured in our experimental setup which is close to Larmor frequency for 1.5T MRI ($63.55$~MHz).

\end{spacing}


%